\documentclass[showpacs,twocolumn,prl,aps,color]{revtex4}
\usepackage{graphicx}
\begin{document}

\title{Orbital Degeneracy and the Microscopic Model of the FeAs Plane in the Iron-Based Superconductors}

\author{Tao Li}
\affiliation{Department of Physics, Renmin University of China,
Beijing, 100872, P.R.China}
\date{\today}

\begin{abstract}
A microscopic model for the FeAs plane in the newly discovered
Iron-based superconductors is proposed with the emphasis on the role
of the orbital degeneracy between the Fe $3d_{xz}$ and $3d_{yz}$
orbital in the crystal field environment. The model predicts
commensurate antiferromagnetic ordered ground state for the parent
compounds and a d-wave superconducting ground state at finite
doping. Inter-orbital couplings plays an interesting role in
enhancing both orders. Correlated variational wave functions are
proposed for both ordered phases.
\end{abstract}

\maketitle

The newly discovered Iron-based superconductors F-doped
LaOFeAs\cite{Kamihara} and its derivatives\cite{Wen,Wang,Chen,Ren}
with rare earth substitution arose many interests in the community.
Similar to the high-T$_{c}$ cuprates, the Iron-based superconductors
also have a layered structure in which the role of the Copper oxygen
plane in the high-T$_{c}$ cuprates is played by the FeAs plane. In
both systems, the charge dopant and the doped carrier are separated
spatially which produces an ideal environment for the coherent
motion of the latter at an incommensurate band filling. This
similarity may imply that a still higher T$_{c}$ is possible in this
series of compounds by optimize the interlayer distance and
planeness of the FeAs layer.

Band structure as calculated from Density functional theory(DFT) and
its DMFT improvement had been examined by several
groups\cite{Haule,Singh,Chao,Fang,Ma}. Not to one's surprise, most
of the spectral weight close to the Fermi energy are contributed by
the Fe 3d orbital, as the Fe is the only ion in the system that
posses a unclosed shell. The DFT calculation also predicted a
antiferromagnetic ordered ground state for the parent compound
LaOFeAs with a ordered moment about 2.3 Bohr magneton, which may
also be consistent with some preliminary experiment results on this
system.

Based on the results of the DFT calculations, several
phenomenological theories for the doped Iron-based superconductors
were proposed\cite{Kuroki,Dai,Han}. In this paper, we follow instead
the quantum chemistry considerations and propose a tight binding
microscopic model for the FeAs plane. In our model, the orbital
degeneracy between the Fe $3d_{xz}$ and $3d_{yz}$ orbital plays an
essential role. Our model predicts commensurate antiferromagnetic
ordered ground state for the parent compound and a d-wave
superconducting ground state at finite doping. Inter-orbital
couplings plays an interesting role in enhancing both kind of
orders. We also propose correlated variational wave functions for
both phases.

A key issue for the construction of a microscopic model for the FeAs
plane is to elucidate the role of the orbital degeneracy of the Fe
3d orbital. In the presence of crystal field, the five fold
degenerate Fe 3d atomic orbital will split according to the
irreducible representations of the crystal symmetry and not all five
3d orbital are equally important for the low energy physics.
According to the DFT calculation, the crystal splitting of the Fe 3d
orbital is small and all five Fe 3d orbital should be included in
the model in principle. However, the inclusion of the electron
correlation effect can enhance the crystal splitting and vice versa.
Here we assume the crystal field splitting to be sufficiently large
for the discussion of the low energy physics and base our discussion
solely on symmetry considerations.

In the FeAs plane, each Fe ion sits at the center of inversion of a
squashed(along the normal of the FeAs plane) tetrahedron formed by
four neighboring As ions(see figure 1). Using the lattice constants
reported in the literature\cite{Haule}, one estimate that the Fe-As
bond make a angle of about 58.8 degree with the normal of the FeAs
plane, while in a perfect tetrahedron the corresponding angle should
be 54.7 degree. In such a local environment of S$_{4}$
symmetry\cite{Chao}, the five fold degenerate 3d orbital split into
three nondegenerate($3d_{3z^{2}-r^{2}}$, $3d_{x^{2}-y^{2}}$, and
$3d_{xy}$) orbital and one doubly degenerate orbital($3d_{xz}$ and
$3d_{yz}$)(see figure 2). If we assume that the crystal field is
contributed mainly by the four neighboring As ions, one would expect
the $3d_{3z^{2}-r^{2}}$ orbital to have the lowest energy, since the
As tetrahedron in squashed in z-direction and the electron cloud of
the $3d_{3z^{2}-r^{2}}$ orbital has the best chance to avoid the As
ion(see figure 1). The next lowest energy orbital would be $3d_{xy}$
whose lobe points toward the vacancy between the As ions in the
plane. If the tetrahedron is not squashed, the $3d_{xy}$ orbital
would be degenerate with the $3d_{3z^{2}-r^{2}}$ orbital. The
squashing distortion of the tetrahedron also lift the degeneracy
between the remaining three orbital, namely
$3d_{x^{2}-y^{2}}$,$3d_{xz}$,$3d_{yz}$. Since the tetrahedron around
the Fe ion is squashed toward the plane, the $3d_{x^{2}-y^{2}}$
orbital should be higher in energy than $3d_{xz}$ and $3d_{yz}$
orbital. In the tetragonal structure of the FeAs plane, the
degeneracy between the $3d_{xz}$ and $3d_{yz}$ orbital is protected
by symmetry.

\begin{figure}[h!]
\includegraphics[width=9cm,angle=0]{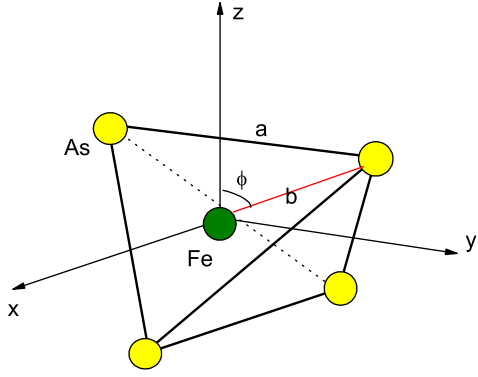}
\caption{The squashed tetrahedron formed by four neighboring As ion
around each Fe ion in the FeAs plane. From the lattice constant
reported in the literature, a=4.020 {\AA}, b=2.35 {\AA}, the angle
$\phi$ between the Fe-As bond with the normal of the Fe-As plane is
estimated to be 58.8 degree, while in a perfect tetrahedron the
angle would be 54.7 degree. Here we use a coordinate system in which
the $x$ and $y$ axis point along the As-As bond directions.}
\label{fig1}
\end{figure}

The divalent Fe ions in the FeAs plane have six electrons in their
3d shell. According to the above splitting scheme, one would expect
a very peculiar situation to occur in the parent compounds in which
the Fermi energy lies within two degenerate bands which are both
half filled. We think this is the reason that the parent compounds
differ from usual band insulator or band metal. It may also hold the
key for the lightly doped LnOFeAs system to show such remarkable
properties as having a T$_{c}$ as high as 50 K\cite{Ren}.

In the following, we assume the crystal splitting to be large enough
to neglect the filled $3d_{3z^{2}-r^{2}}$  and $3d_{xy}$ band and
the empty $3d_{x^{2}-y^{2}}$ band in the discussion of low energy
physics. With this simplification, we are left with a model with two
degenerate bands. In this model, each Fe site can accommodate at
most four electrons in the two degenerate bands. In the parent
compound, each Fe site have two electrons on average in the two
degenerate bands and the system is thus half filled. The
interactions between the electrons on the same Fe site can be
classified into three types, namely the intra-orbital Hubbard
repulsion U , the spin independent part of the inter-orbital Hubbard
repulsion U$_{1}$, and the spin dependent part of the inter-orbital
Hubbard repulsion J(the usual Hund's rule coupling). From the
definition of these terms, one easily see that that the inequality
$U>U_{1}>\frac{J}{4}>0$ should be satisfied. These interaction terms
is represented by the model Hamiltonian
\begin{eqnarray}
 H_{U}=U\sum_{i}(n_{i,xz,\uparrow}n_{i,xz,\downarrow}+n_{i,yz,\uparrow}n_{i,yz,\downarrow}) \nonumber \\
 +U_{1}\sum_{i}(n_{i,xz,\uparrow}+n_{i,xz,\downarrow})(n_{i,yz,\uparrow}+n_{i,yz,\downarrow}) \nonumber \\
 -J\sum_{i} \vec{S}_{i,xz} \cdot \vec{S}_{i,yz}
\end{eqnarray}
in which
$n_{i,xz,\uparrow}=c^{\dagger}_{i,xz,\uparrow}c_{i,xz,\uparrow}$
denotes the number of up spin electron in the $3d_{xz}$ orbital.
$\vec{S}_{i,xz}=\frac{1}{2}c^{\dagger}_{i,xz,\alpha}\vec{\sigma}_{\alpha,\beta}c_{i,xz,\beta}$
denotes the spin density on the $3d_{xz}$ orbital. The meaning of
other terms in the equation is similar.

\begin{figure}[h!]
\includegraphics[width=9cm,angle=0]{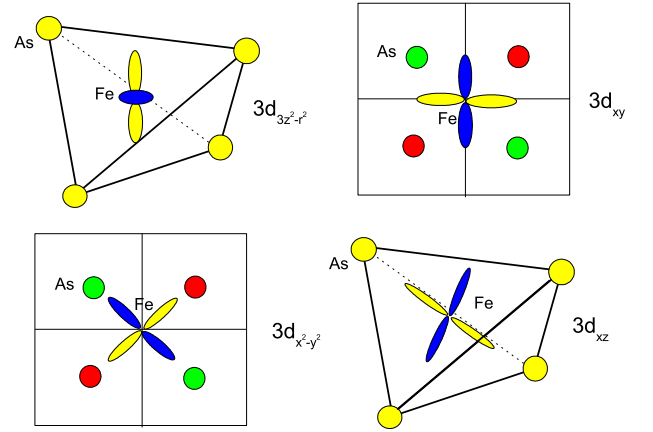}
\caption{The five d-orbital in the crystal field of the tetrahedron
formed by the four neighboring As ions around each Fe ion. The
$3d_{x^{2}-y^{2}}$ and the $3d_{xy}$ orbital are shown here in a
projection onto the $xy$ plane. In the projection graph, red(green)
filled circles denote As ion above(below) the $xy$ plane. Here we
only show the $3d_{xz}$ orbital, the $3d_{yz}$ orbital can be
obtained from the $3d_{xz}$ orbital through a rotation of
$\frac{\pi}{2}$ along the z-axis.} \label{fig2}
\end{figure}

In the FeAs plane, the Fe ions are too remote from each other to
have direct hoping and the As ion is needed to bridge the hoping
between the Fe ions. The As ion has three 3p orbital in its outmost
shell. Since the $3p_{z}$ orbital is far away from the Fermi energy,
we consider only the hopping path mediated by the $3p_{x}$ and
$3p_{y}$ orbital of As ion. Symmetry considerations shows that the
hoping mediated by the As $3p_{x}$ and $3p_{y}$ orbital conserve the
orbital index, namely, an electron initially on the Fe
$3d_{xz}(3d_{yz})$ orbital can only hop onto the $3d_{xz}(3d_{yz})$
orbital of destination Fe ion(the inclusion of the hopping path
mediated by $3p_{z}$ orbital will break such a symmetry)(see figure
3). Another peculiarity about the As ion aided effective hoping is
that the hoping integral between next-nearest-neighboring Fe ions is
anisotropic(the hoping between nearest neighboring Fe ions remains
isotropic). This anisotropy depends on the orbital and is a
reflection of the anisotropic nature of the $3d_{xz}$ and $3d_{yz}$
orbital. Taking all these considerations into account, we arrive at
the following model for the effective hoping between the Fe ions.
\begin{eqnarray}
 H_{t}=-t\sum_{i,\vec{\delta},\sigma}\left[(c^{\dagger}_{i,xz,\sigma}c_{i+\vec{\delta},xz,\sigma}+c^{\dagger}_{i,yz,\sigma}c_{i+\vec{\delta},yz,\sigma})+H.C. \right] \nonumber \\
 -t_{1}\sum_{i,\sigma}\left[(c^{\dagger}_{i,xz,\sigma}c_{i+\vec{\delta}^{'},xz,\sigma}+c^{\dagger}_{i,yz,\sigma}c_{i+\vec{\delta}^{''},yz,\sigma})+H.C.\right] \nonumber \\
 -t_{2}\sum_{i,\sigma}\left[(c^{\dagger}_{i,xz,\sigma}c_{i+\vec{\delta}^{''},xz,\sigma}+c^{\dagger}_{i,yz,\sigma}c_{i+\vec{\delta}^{'},yz,\sigma})+H.C.\right],
\end{eqnarray}
in which $\vec{\delta}=\vec{e_{x}}$ or $\vec{e_{x}}$ denotes the
nearest neighboring vectors on the square lattice form by Fe ions.
$\vec{\delta}^{'}=\vec{e_{x}}+\vec{e_{y}}$ and
$\vec{\delta}^{''}=\vec{e_{x}}-\vec{e_{y}}$ denotes the next nearest
neighboring vectors. Since there are two hoping paths between the
nearest neighboring Fe ions but only one for the next-nearest
neighboring Fe ions, $t$ is roughly a factor of two larger than both
$t_{1}$ and $t_{2}$. From symmetry considerations, one can also show
that all of the three hoping integrals, $t$, $t_{1}$ and $t_{2}$ can
be taken positive with a suitable choice of gauge. The dispersion
relation of the $3d_{xz}$ band is given by
\begin{eqnarray}
 E_{xz,\mathrm{k}}=-2t(\cos(k_{x})+\cos(k_{x}))-2t_{1}\cos(k_{x}+k_{y})\nonumber \\
-2t_{2}\cos(k_{x}-k_{y}),
\end{eqnarray}
the dispersion relation of the $3d_{yz}$ band can be obtained from
that of $3d_{zz}$ band by the reflection in momentum space $k_{x}
\rightarrow -k_{x}$.

The total Hamiltonian of the system is the sum of the on-site term
$H_{U}$ and the hopping term $H_{t}$. Before looking at the phase
diagram of the model, we first estimate the model parameters.
According to DFT calculations, the Hubbard repulsion $U$  and
$U_{1}$ are about $4 eV$ in magnitude and the Hund's rule coupling
$J$ is about $1 eV$ in magnitude. For the hopping terms, as both the
$\mathrm{Fe} 3d_{xz} - \mathrm{As} 3p_{x}$ separation and the hoping
integral between the two are of the order of $1 eV$, we estimate the
effective hoping integral between the Fe ions mediated by the As
ions to be also of the order of $1 eV$. Thus the system has a
moderate level of electron correlation and moderate level of
frustration(as indicated by the ratio between $t$ and $t_{1}$).
\begin{figure}[h!]
\includegraphics[width=9cm,angle=0]{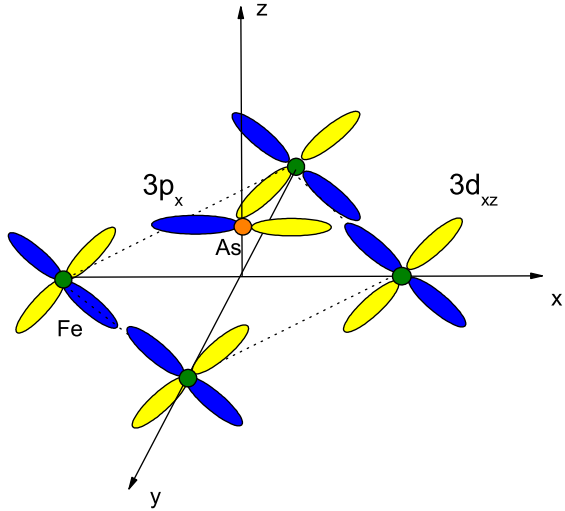}
\caption{A schematic representation of the As $3p$ orbital aided
hoping between neighboring Fe ions on the square lattice. Note that
the As ions lies above or below the Fe plane. Simple symmetry
consideration shows that the electron on Fe $3d_{xz}$($3d_{yz}$)
orbital can hop only through the As $3p_{x}$($3p_{y}$) orbital. This
explains the conservation of orbital index and the anisotropy of the
hoping Hamiltonian. The electron wave function on the yellow(blue)
lobes have positive(negative) values. } \label{fig3}
\end{figure}

Now we discuss the possible phase diagram of the model. In the half
filled parent compounds, each site is occupied on average by two
electrons. Since the intra-orbital Hubbard repulsion $U$ is larger
than the inter-orbital Hubbard repulsion $U_{1}$, the two electrons
tend to occupy different orbital and the remaining Hund's rule
coupling tends to align up the spin of the electrons on both
orbital. Thus each individual Fe ion carries approximately a unit
spin angular momentum and shows two Bohr magneton of magnetic
moment. However, the inclusion of the hoping terms between the Fe
ions will induce antiferromagnetic spin exchange between neighboring
Fe ions within each band. Since the frustration is moderate, one can
expect an commensurate antiferromagnetic order to appear in the
ground state of parent compound. As a result of the Hund's rule
coupling, the antiferromagnetic order in the $3d_{xz}$ and the
$3d_{yz}$ band enhance with each other. With this understanding in
mind, we can write down the following variational wave function for
the ground state of the half filled system,
\begin{equation}
 |\Psi\rangle=|\mathrm{SDW}\rangle_{xz} \otimes |\mathrm{SDW}\rangle_{yz},
\end{equation}
in which $|\mathrm{SDW}\rangle_{xz}$ and $|\mathrm{SDW}\rangle_{yz}$
denote the SDW ordered state in the $3d_{xz}$ and the $3d_{yz}$
band.

In our model, the two bands are correlated with each other through
the inter-orbital Hubbard term and the Hund's rule coupling. To
describe such correlation effect, we introduce the following
Jastrow-type variational wave function
\begin{equation}
 |\Psi\rangle=g^{\hat{D}}g_{1}^{\hat{V}}g_{J}^{\hat{S}}|\mathrm{SDW}\rangle_{xz} \otimes |\mathrm{SDW}\rangle_{yz},
\end{equation}
in which $0<g,g_{1},g_{J}<1$ are Gutzwiller factors introduced to
describe the correlation effect caused by intra- and inter-orbital
Hubbard resulpsion and Hund's rule coupling. The operators $\hat{D},
\hat{V}$ and $\hat{S}$ are given by
$\hat{D}=\sum_{i}(n_{i,xz,\uparrow}n_{i,xz,\downarrow}+n_{i,yz,\uparrow}n_{i,yz,\downarrow})$,
$\hat{V}=\sum_{i}(n_{i,xz,\uparrow}+n_{i,xz,\downarrow})(n_{i,yz,\uparrow}+n_{i,yz,\downarrow})$
and $\hat{S}=\sum_{i}(\vec{S}_{i,xz} \cdot
\vec{S}_{i,yz}+\frac{3}{4})$. In the antiferromagnetic ordered
ground state, we expect the correction due to the correlation effect
to be small as the main effect of the inter-band correlation is
embodied in the mutual enhancement of the AF order in the two bands.

At finite doping, the antiferromagnetic exchange within each band
will induce pairing instability in the d-wave channel as in the
high-T$_{c}$ problem. Unlike the high-T$_{c}$ problem, now we have
two degenerate bands which are correlated with each other by the
inter-orbital Hubbard repulsion and the Hund's rule coupling. As in
the antiferromagnetic ordered state, it is likely that the
inter-orbital coupling will induce mutual enhancement of the pairing
instability in both bands. However, the inter-orbital coupling will
inevitably increase the effective mass of the Cooper pair in each
band and thus reduce the superfluid density. With this understanding
in mind, we can write down a correlated variational wave function
for the superconducting state as we have done for the
antiferromagnetic ordered state.
\begin{equation}
 |\Psi\rangle=g^{\hat{D}}g_{1}^{\hat{V}}g_{J}^{\hat{S}}|\mathrm{d-BCS}\rangle_{xz} \otimes |\mathrm{d-BCS}\rangle_{yz},
\end{equation}
in which $|\mathrm{d-BCS}\rangle_{xz}$ and
$|\mathrm{d-BCS}\rangle_{yz}$ denote the d-wave BCS superconducting
state in the $3d_{xz}$ and the $3d_{yz}$ band. We note that the
Gutzwiller factor describing the Hund's rule coupling,
$g_{J}^{\hat{S}}$, now plays a very similar role as that of the
projection operator in the construction of the matrix product type
ground state of the AKLT model\cite{Affleck}. Thus the
superconducting state can be more appropriately called a condensate
of entangled Cooper pairs rather than independent Cooper pairs.

The arguments presented above are rather crude and a more careful
study of the model proposed in this paper is obviously needed to
elucidate its relationship with the superconductivity observed in
F-doped LnOFeAs. In principle, the arguments can be made more
quantitative through systematic mean field treatment or Gutzwiller
approximation. A direct numerical calculation on the variational
wave function is also possible. However, before any such effort is
made, it is important first to know the actual value of the crystal
field splitting by, for example, the optical measurement, and check
the validity of neglecting the $3d_{3z^{2}-r^{2}}$,
$3d_{x^{2}-y^{2}}$ and the $3d_{xy}$ orbital in the low energy
subspace.

In summary, a microscopic model for the FeAs plane of the newly
discovered Iron-based superconductors is proposed. In our model, the
orbital degeneracy between the Fe $3d_{xz}$ and the $3d_{yz}$
orbital plays an important role and may explain the peculiarities of
the parent compounds of this family of new superconductors. The
model predicts a commensurate antiferromagnetic ordered ground state
for the parent compound and a d-wave pairing superconducting ground
state in the doped case. In the latter case, the inter-orbital
coupling transform the system from a condensate of independent
Cooper pairs into a condensate of entangled Coopers and it would be
interesting to see effect of inter-orbital couplings on the
dynamical properties of the new superconductor.

This work is supported by NSFC Grant No. 10774187. The author
acknowledge the discussion with Prof. Zheng-Yu Weng, Prof. Ya-Yu
Wang, Prof. Yue-Hua Su and Kai Wu.

\bigskip

\end{document}